\documentstyle[prd,aps,epsfig]{revtex}
\begin{document}
\title{Weak annihilation in the rare radiative $B\to \rho\gamma$ decay}
\author{M. Beyer,$^a$ 
D. Melikhov,$^{b,c}$
N. Nikitin, $^c$ and 
B. Stech$^b$}
\address{$^a$ Fachbereich Physik, Universit\"at Rostock, 
Universit\"atsplatz, 3, D-18051, Rostock, Germany \\
$^b$ Institut f\"ur Theoretische Physik, Universit\"at Heidelberg,  
Philosophenweg 16,  D-69120, Heidelberg, Germany \\
$^c$ 
Institute of Nuclear Physics, Moscow State University, 119899, Moscow, Russia}
\maketitle
\begin{abstract}
The ampitude of the $B\to\rho\gamma$ decay induced by the flavour-changing neutral currents 
contains the penguin contribution and the weak-annihilation contribution generated by the 
4-quark operators in the effective Hamiltonian. The penguin contribution is known quite well. 
We analyze the weak-annihilation which is suppressed by the heavy-quark mass compared 
to the penguin contribution. 

In the factorization approximation, the weak annihilation amplitude is represented in terms of the leptonic 
decay constants and the meson-photon matrix elements of the weak currents. The latter contain 
the $B\gamma$, $\rho\gamma$ transition form factors and contact terms determined by the 
equations of motion. We calculate the $B\gamma$ and $\rho\gamma$ form factors within the 
relativistic dispersion approach and obtain numerical estimates for the weak annihilation amplitude.  
\end{abstract}

\section{Introduction}
The investigation of rare semileptonic $B$ decays induced by the 
flavour-changing neutral current transitions $b\to s$ and $b\to d$ represents an important 
test of the Standard Model or its extentions. 
Rare decays are forbidden at the tree level in the Standard Model and occur through 
loop diagrams. Thus they provide the possibility to probe the structure of the 
electroweak sector at large mass scales from contributions of virtual particles in the loop
diagrams. Interesting
information about the structure of the theory is contained in the Wilson coefficents in the 
effective Hamiltonian which describes the $b\to s,d$ transition at low energies. These Wilson 
coefficients take different values in different theories with testable 
consequences in rare $B$ decays. 

Among rare $B$ decays the radiative decays $b\to s\gamma$ and $b\to d\gamma$ have the 
largest probabilites. The $b\to s\gamma$ transitions are CKM favoured and have larger 
branching ratios than the $b\to d\gamma$ transitions.  
The $b\to s\gamma$ transition has been observed by CLEO in the exclusive channel    
$B\to K^*\gamma$ in 1993 and measured inclusively in 1995. 
The $B\to \rho\gamma$ decay will be extensively studied by BaBar and BELLE. 

The main uncertainty in the theoretical analysis of $B$ decays is connected with  
long-distance QCD effects arising from the presence of hadrons in initial and 
final states. In inclusive decays these effects are under better control, 
however from inclusive measurements it is more difficult to obtain precise results. 

The decay amplitude contains two different contributions: 
one arising from the electromagnetic penguin operator and another from the 4-fermion 
operators in the effective Hamiltonian. One of the effects generated by the 
4-fermion operators is the weak annihilation (WA).  
Further details about short- and long-distance effects in the radiative decays 
can be found in recent publications \cite{gp,m2001,ap,ppp} and references therein. 

In the $B\to K^*\gamma$ decay the weak annihilation is negligible compared to the 
penguin effect: it is suppressed by two powers of the small parameter $\lambda\simeq 0.2$ 
of the Cabbibo-Kabayashi-Maskawa (CKM) matrix. In $B\to \rho\gamma$ 
both effects have the same order in $\lambda$ and must be taken into account. 

The penguin contribution has been analyzed in several ways and is known quite well. 
On the other hand, the WA in $B\to \rho\gamma$ has been studied in less detail: the relevant 
form factors were analyzed within sum rules \cite{sr1,sr2} 
and perturbative QCD \cite{korch}. However, some contributions to these form factors were 
neglected. These contributions may be relevant if precise measurements become available. 

The aim of this paper is to analyze the weak annihilation for the $B^-\to \rho^-\gamma$ 
decay more closely. 

\begin{center}
\begin{figure}[hb]
\mbox{\epsfig{file=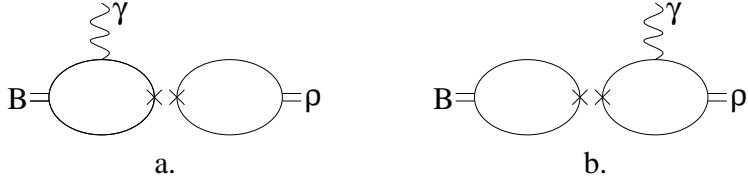,width=10cm}}
\caption{\label{fig:diag}
Diagrams describing the weak annihilation process for $B\to\rho\gamma$ in the factorization approximation: 
(a) The photon is emitted from the loop containing the $b$ quark, 
(b) The photon is emitted from the loop containing only light quarks.}
\end{figure}
\end{center}

In the factorization approximation, the weak-annihilation amplitude can be represented as the product 
of meson leptonic decay constants and matrix elements of the weak current between 
meson and photon. The latter contain the meson-photon transition form factors and 
contact terms which are determined by the equations of motion. The photon can be emitted from the 
loop containing the $b$ quark which is described by $B\gamma$ transition form factors. It can also 
be emitted from the loop containing only light quarks described by the $\rho\gamma$ 
transition form factors (Fig \ref{fig:diag}).  

We calculate the $B\gamma$ form factors within the relativistic dispersion approach which 
expresses these form factors in terms of the $B$ meson wave function. 
We demonstrate that the form factors calculated by the dispersion approach 
behave in the limit $m_b\to\infty$ in agreement with perturbative QCD. 
The $B$ meson wave function was previously tested in the $B\to$ light meson weak decays and 
is known quite well, allowing us to provide reliable numerical estimates for the $B\gamma$ form factors.    

The $\rho\gamma$ transition form factor is related to the divergence of the vector and 
axial-vector currents. In the case of the axial-vector current it is   
proportional to the light-quark masses if the classical equation of motion is applied. 
Because the quark momenta in the loop are high, these masses have to be identified with 
current quark masses. For this reason the corresponding $\rho\gamma$ form factor was neglected in previous 
analyses \cite{gp,sr1}.    
We find however that this argument is not correct: this form factor 
remains finite in the limit of vanishing light quark mass $m\to 0$ and behaves like 
$\sim M_\rho f_\rho/M_B^2$ which means the violation of the classical equation of motion.  
We present the result for the $\rho\gamma$ transition form factor but leave a detailed 
discussion 
of the anomaly appearing for the matrix element 
$\langle \rho\gamma|\partial_\nu A_\nu|0\rangle$ for a special publication. 

Finally, we provide numerical estimates of the weak-annihilation amplitude 
taking into account the $B\gamma$, $\rho\gamma$ transition form factors, and the 
contact term contributions.   

In Section II the effective Hamiltonian for the $b\to d$ transition and the general 
structure of the amplitude are presented. In Section III we discuss the photon emission from the 
$B$ meson loop and obtain the $B\gamma$ transition form factors within the dispersion 
approach.  
Section IV contains results for the $\rho\gamma$ transition form factors.  
In Section V the numerical estimates are given. The concluding Section 
summarises our results. 

\section{The effective Hamiltonian, the amplitude and the decay rate}
The amplitude of the weak radiative 
$B\to \rho$ transition is given by the matrix element of the effective Hamiltonian 
for the $b\to d$ transition 
\begin{eqnarray}
A(B\to\rho\gamma)=\langle \gamma(q_1)\rho(q_2)|H_{\rm eff}(b\to d)|B(p) \rangle,  
\end{eqnarray}
where $p$ is the $B$ momentum, $q_2$ is the $\rho$ momentum, and $q_1$ is the photon
momentum, $p=q_1+q_2$, $q_1^2=0$, $q_2^2=M_\rho^2$, $p^2=M_B^2$. 
The effective weak Hamiltonian has the structure\cite{gsw}:
\begin{eqnarray}
\label{Heff}
H_{\rm eff}(b\to d) &=& \frac{G_F}{\sqrt{2}}\,\xi_t
C_{7\gamma}(\mu){\cal O}_{7\gamma}
-\frac{G_F}{\sqrt{2}}{\xi_u}
\left (C_1(\mu){\cal O}_1+C_2(\mu){\cal O}_2\right), 
\end{eqnarray}
where only operators relevant for the penguin and weak ahnnihilation effects are listed. 
$G_F$ is the Fermi constant, $\xi_q=V^*_{qd}V_{qb}$, $C_i$'s are the Wilson 
coefficients and ${\cal O}_i$'s are the basis operators 
\begin{eqnarray}
{\cal O}_{7\gamma} &=& \frac{e}{8\pi^2}\,
\bar d_{\alpha}\sigma_{\mu\nu}
m_b(\mu)\left (1+\gamma_5\right )\, b_{\alpha}\, F_{\mu\nu}, \\
{\cal O}_1 &=& \bar d_{\alpha}\gamma_{\nu}(1-\gamma_5)u_{\alpha}\;
\bar u_{\beta}\gamma_{\nu}(1-\gamma_5) b_{\beta},\nonumber
\\
{\cal O}_2 &=& \bar d_{\alpha}\gamma_{\nu}(1-\gamma_5) u_{\beta}\; 
\bar u_{\beta}\gamma_{\nu}(1-\gamma_5) b_{\alpha},\nonumber\\
\end{eqnarray}
with the following notation: $e=\sqrt{4\pi\alpha_{\rm em}}$, 
$\gamma^5=i\gamma^0\gamma^1\gamma^2\gamma^3$,  
$\sigma_{\mu\nu}=i\left [\gamma_{\mu},\gamma_{\nu}\right ]/2$,
$\epsilon^{0123}=-1$ and
${\rm Sp}\left (\gamma^5\gamma^{\mu}\gamma^{\nu}\gamma^{\alpha}\gamma^{\beta}\right )
=4i\epsilon^{\mu\nu\alpha\beta}$, $F_{\mu\nu}={\partial_\mu A_\nu-\partial_\nu A_\mu}$.
 
The amplitude can be parametrized as follows  
\begin{eqnarray}
A(B^-\to\rho^-\gamma)=\frac{eG_F}{\sqrt{2}}
\left[
\epsilon_{q_1\epsilon^\ast_1 q_2 \epsilon_2^\ast}F_{\rm PC}
+i \epsilon_2^{\ast\nu}\epsilon_1^{\ast\mu} \left(g_{\nu\mu}\,pq_1-p_\mu q_{1\nu}\right)F_{\rm PV}
\right], 
\end{eqnarray}
where $F_{\rm PC}$ and $F_{\rm PV}$ are the parity-conserving and 
parity-violating invariant amplitudes, respectively. 
$\epsilon_2$($\epsilon_1$) is the 
$\rho$-meson (photon) polarization vector. We use the short-hand notation 
$\epsilon_{abcd}=\epsilon_{\alpha\beta\mu\nu}a^{\alpha}b^{\beta}c^{\mu}d^{\nu}$ 
for any 4-vectors $a,b,c,d$. 

For the decay rate one finds 
\begin{eqnarray}
\label{rate}
\Gamma(B^-\to\rho^-\gamma)=\frac{G^2_F\,\alpha_{em}}{16}M_B^3
\left(1-{M^2_\rho}/{M_B^2}\right)^3
     \left( |F_{\rm PC}|^2+|F_{\rm PV}|^2 \right). 
\end{eqnarray}

\subsection{The penguin amplitude}
The main contribution to the amplitude is given by the  
electromagnetic penguin operator $O_{7\gamma}$: 
\begin{eqnarray}
A_{\rm peng}(B\to\rho\gamma)&=&
-\frac{eG_F}{\sqrt{2}}\xi_t C_7\frac{m_b}{2\pi^2} T_1(0)
\left(
\epsilon_{q_1\epsilon_1^* q_2 \epsilon_2^*}
+i\epsilon_2^{\ast\nu}\epsilon_1^{\ast\mu}(g_{\nu\mu}\,pq_1-p_\mu q_{1\nu})
\right),
\end{eqnarray}
where $T_1$ is the form factor of the $B\to\rho$ transition through 
the tensor current \cite{wsb,m,ms}.
The corresponding contribution to the invariant amplitudes is therefore 
\begin{eqnarray}
\label{fpeng}
F^{\rm peng}_{\rm PC}=F^{\rm peng}_{\rm PV}=-\xi_t C_7\frac{m_b}{2\pi^2}T_1(0). 
\end{eqnarray}

\subsection{The weak annihilation amplitude}
The radiative $B\to\rho \gamma$ transition also receives contribution from the 
4-fermion operators ${\cal O}_1$ and ${\cal O}_2$.  
For the charged $B^-\to\rho^-(q_2)\gamma(q_1)$ transition the corresponding amplitude reads 
\begin{eqnarray}
\label{WA}
A_{\rm WA}(B^-\to\rho^-\gamma)&=&-\frac{G_F}{\sqrt{2}}\xi_u 
\langle \rho(q_2)\gamma(q_1)|\bar d\gamma_\nu(1-\gamma_5)u\cdot 
\bar u\gamma_\nu(1-\gamma_5)b|B(p)\rangle,  
\end{eqnarray}  
In what follows we suppress the lable WA.   
Neglecting the nonfactorizable soft-gluon exchanges, i.e. assuming vacuum saturation, 
the complicated matrix element in Eq. (\ref{WA}) is reduced to simpler quantities - 
the meson-photon matrix elements of the bilinear quark currents and the 
meson decay constants. The latter are defined as usual  
\begin{eqnarray}
\langle \rho(q_2)|\bar d\gamma_\nu u|0\rangle &=& \epsilon_{2\nu}^\ast M_\rho f_\rho, \qquad  f_\rho>0,
\nonumber \\
\langle 0|\bar u\gamma_\nu \gamma_5 b|B(p)\rangle &=& ip_\nu f_B,\qquad  f_B>0.
\end{eqnarray}
It is convenient to isolate the parity-conserving contribution which emerges from the
product of the two equal-parity currents, and the parity-violating contribution 
which emerges from the product of the two opposite-parity currents. 

\subsubsection{The parity-violating amplitude} 
The parity-violating amplitude has the form  
\begin{eqnarray}
\label{apva}
A_{\rm PV}(B\to\rho\gamma)&=&\frac{G_F}{\sqrt{2}}\xi_u a_1  
\left\{
\langle \rho\gamma|\bar d \gamma_\nu u|0 \rangle 
\langle 0|\bar u \gamma_\nu\gamma_5 b|B \rangle  
+
\langle \rho|\bar d \gamma_\nu u|0 \rangle 
\langle \gamma|\bar u \gamma_\nu\gamma_5 b|B \rangle \right\}. 
\end{eqnarray}
Here $a_1$ is an effective Wilson coefficient, which we take as $a_1=C_1+C_2/N_c$ 
at the scale $\sim$ 5 GeV.  

The first term is a contact term which can be calculated using the equations 
of motion for the quark fields \cite{gp}. 
Setting $m_u=m_d$, we find 
\begin{eqnarray}
\label{term1}
\langle \rho\gamma|\bar d \gamma_\nu u|0 \rangle 
\langle 0|\bar u \gamma_\nu\gamma_5 b|B \rangle=
ip_\nu f_B \langle \rho\gamma|\bar d \gamma_\nu u|0 \rangle =
f_B \langle \rho\gamma|\partial_\nu(\bar d \gamma_\nu u)|0 \rangle =
ie\epsilon_2^{\ast\nu}\epsilon_1^{\ast\mu} \;g_{\mu\nu}M_\rho f_\rho f_B. 
\end{eqnarray}
The $B\to\gamma$ amplitude in the second term of (\ref{apva})
can be parametrized as follows 
\begin{eqnarray}
\label{term2}
\langle \gamma(q_1)|\bar u \gamma_\nu\gamma_5 b|B(p) \rangle=
ie\epsilon_1^{\ast\mu} \, \left[
(g_{\nu\mu}\,pq_1-p_\mu q_{1\nu})\frac{2F_A}{M_B}
-p_\mu q_{1\nu} \frac{2f_B}{M_B^2-M_\rho^2}\right].    
\end{eqnarray}
It contains the form factor $F_A(q_2^2=M_\rho^2)$, and the contact term proportional to $f_B$ 
(the derivation of this relation is given in the Appendix). 

Summing the contributions of the photon emission from the $B$-meson loop and the $\rho$-meson 
loop gives the amplitude $A_{\rm PV}$ which can be represented in the form 
$A_{\rm PV}=\epsilon_1^{\ast\mu} T_\mu$ with  
$q_1^\mu T_\mu=0$ as required by gauge invariance. 
Thus, the weak-annihilation contribution to the form factor $F_{\rm PV}$ 
for the $B^-\to\rho^-\gamma$ decay is\footnote{We note that for the $B^+\to\rho^+\gamma$ 
the term proportional to $f_B$ changes sign (see Appendix), and also 
$F_A^{B^+}=-F_A^{B^-}$ as can be obtained from charge conjugation of 
the amplitude $A_{\rm PV}$. For the $B^0\to\rho^0\gamma$ decay the term $\sim f_B$ is absent.}
\begin{eqnarray}
\label{fpv}
F^{\rm WA}_{\rm PV}=\xi_u a_1 f_\rho M_\rho\;\left[\frac{2F_{A}}{M_B}+\frac{2f_B}{M_B^2-M_\rho^2}\right].
\end{eqnarray}
The two contact terms which are present in the amplitudes of the photon
emission from the $B$ meson loop and from the $\rho$-meson loop do not cancel each other
(we disagree here with the claim of Ref. \cite{sr2}) but lead to a nonvanishing contribution proportional to $f_B$. 

\subsection{The parity-conserving amplitude} 
This amplitude reads  
\begin{eqnarray}
\label{apc}
A_{\rm PC}(B\to\rho\gamma)&=&-\frac{G_F}{\sqrt{2}}\xi_u a_1 
\left\{\langle \rho|\bar d\gamma_\nu u|0 \rangle \langle \gamma|\bar u \gamma_\nu b|B \rangle 
+
\langle \gamma\rho|\bar d\gamma_\nu \gamma_5 u |0 \rangle 
\langle 0| \bar u\gamma_\nu \gamma_5 b|B \rangle \right\}. 
\end{eqnarray}
The $B\to\gamma$ amplitude from the first term in the brackets contains the form factor
$F_V(q_2^2=M_\rho^2)$:   
\begin{eqnarray}
\langle \gamma(q_1)|\bar u\gamma_\nu b|B(p)\rangle = 
e\,\epsilon_{\nu\mu p q_1} \epsilon_1^{\ast\mu} \;\frac{2F_V}{M_B}.
\end{eqnarray}
The second term in (\ref{apc}) can be reduced to the divergence of the 
axial-vector current and contains another form factor, $G_V$: namely,   
\begin{eqnarray}
\langle 0|\bar u\gamma_\nu \gamma_5 b|B\rangle 
\langle \gamma\rho|\bar d\gamma_\nu\gamma_5 u|0 \rangle=
f_B \langle \gamma\rho|\partial_\nu \bar d\gamma_\nu\gamma_5 u |0 \rangle=
e\,f_B G_V \epsilon_{q_1\epsilon_1^\ast q_2\epsilon_2^\ast}. 
\end{eqnarray}
Thus, the weak annihilation contribution to $F_{\rm PC}$ reads  
\begin{eqnarray}
\label{fpc}
F^{\rm WA}_{\rm PC}=\xi_u a_1 M_\rho f_\rho\left[\frac{2F_V}{M_B}-\frac{f_B\;G_V}{M_\rho f_\rho}\right]. 
\end{eqnarray}
Summing up, within the factorization approximation the weak annihilation amplitude 
can be expressed in terms of the three form factors $F_A$, $F_V$, and $G_V$. 

\section{The form factors $F_A$ and $F_V$}
In this section we derive the formulas for the form factors $F_{A,V}$ within the dispersion approach 
to the transition form factors. This approach has been formulated in detail in 
\cite{m} and applied to the weak decays of heavy mesons in \cite{ms}. Recall that the 
form factors $F_{A,V}$ describe the transition of the $B$-meson to the photon with the momentum 
$q_1$, $q_1^2=0$, induced by the axial-vector (vector) current with the momentum transfer 
$q_2$, $q_2^2=M_\rho^2$. For technical reasons, it is convenient to treat the form factor $F_{A(V)}$ as describing the 
amplitude of the photon-induced transition of the $B$-meson into a $b\bar u$ axial-vector 
(vector) virtual particle with the corresponding factor $1/(s-q_2^2)$ in the dispersion integral. 
Then we can directly apply the equations obtained in \cite{m} for the 
meson-meson transition form factors. 

\subsection{The form factor $F_A$}
\begin{center}
\begin{figure}[hb]
\mbox{\epsfig{file=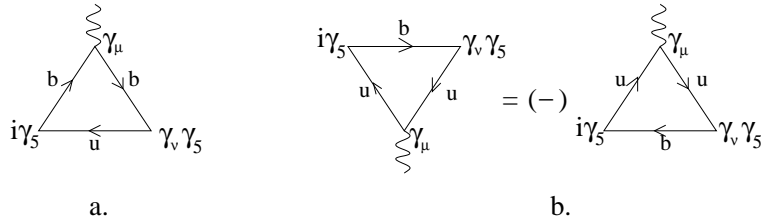,width=10cm}}
\caption{\label{fig:Fa}
Diagrams for the form factor $F_A$: a) $F_A^{(b)}$, b) $F_A^{(u)}$.}
\end{figure}
\end{center}

The form factor $F_A$ is given by the diagrams of Fig \ref{fig:Fa}. 
Fig \ref{fig:Fa}a shows $F_A^{(b)}$, 
the contribution to the form factor of the process when the $b$ quark interacts with the 
photon; Fig \ref{fig:Fa}b describes the contribution of the process when the quark $u$ interacts while 
$b$ remains a spectator. One has    
\begin{eqnarray}
F_A=F_A^{(b)}+F_A^{(u)}. 
\end{eqnarray} 
The $B^-$ meson is described 
by the vertex $\bar b(k_b)\; i\gamma_5 u(k_u)\;G(s)/{\sqrt{N_c}}$, with 
$G(s)=\phi_B(s)(s-M_B^2)$. The $B$-meson wave function $\phi_B$ is normalized according to the relation \cite{m}
\begin{eqnarray}
\label{norma}
\frac{1}{8\pi^2}\int\limits_{(m_b+m_u)^2}^\infty ds \phi_B^2(s)
\left({s-(m_b-m_u)^2}\right)\frac{\lambda^{1/2}(s,m_b^2,m_u^2)}{s}=1.  
\end{eqnarray}
Here $\lambda(a,b,c)=(a-b-c)^2-4bc$ is the triangle function. 

It is convenient to change the direction of the quark line in the loop diagram of 
Fig \ref{fig:Fa}b. This is done by performing the charge conjugation of the matrix element 
and leads to a sign change for the $\gamma_\nu\gamma_5$ vertex. 

\begin{center}
\begin{figure}[hb]
\mbox{\epsfig{file=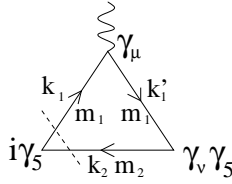,width=3cm}}
\caption{\label{fig:Fat}
The triangle diagram for $F_A^{(1)}(m_1,m_2)$. The cut corresponds to calculating the 
imaginary part in the variable $p^2$. }
\end{figure}
\end{center}

Now both diagrams in Fig \ref{fig:Fa} a,b are reduced to the diagram of Fig \ref{fig:Fat} 
which defines the 
form factor $F_A^{(1)}(m_1,m_2)$: Setting $m_1=m_b$, $m_2=m_u$ gives $F_A^{(b)}$: 
$F_A^{(b)}=Q_b F_A^{(1)}(m_b,m_u)$. Similarly, setting $m_1=m_u$, $m_2=m_b$ 
gives $F_A^{(u)}$, $F_A^{(u)}=-Q_u F_A^{(1)}(m_u,m_b)$. 
For the diagram of Fig \ref{fig:Fat} (quark 1 emits the photon, quark 2 is the spectator)    
the trace reads  
\begin{eqnarray}
\nonumber 
-{\rm Sp}\;i\gamma_5(m_2-\hat k_2)\gamma_\nu\gamma_5(m_1+\hat k'_1)\gamma_\mu(m_1+\hat k_1)=
4i(k_1+k_1')_\mu(m_1 k_2+m_2 k_1)_\nu+
4i(g_{\mu\nu}q_\alpha-g_{\mu\alpha}q_\nu)(m_1 k_2+m_2 k_1)_\alpha. 
\end{eqnarray}
The spectral density of the form factor $F^{(1)}_A(m_1,m_2)$ in the variable $p^2$, $p=k_1+k_2$,  
is the coefficient of the structure $g_{\mu\nu}$ obtained after the integration of the trace over 
the quark phase space. Performing necessary calculations, we arrive at the following single dispersion integral  
\begin{eqnarray}
\label{fadisp}
\frac{2}{M_B}F_{A}^{(1)}&=&\frac{\sqrt{N_c}}{4\pi^2}\int\limits_{(m_b+m_u)^2}^\infty
\frac{ds\;\phi_B(s)}{(s-M_\rho^2)}\left\{
\left(m_1\log\left(\frac{s+m_1^2-m_2^2+\lambda^{1/2}(s,m_1^2,m_2^2)}
{s+m_1^2-m_2^2-\lambda^{1/2}(s,m_1^2,m_2^2)}\right)
+(m_2-m_1)\frac{\lambda^{1/2}(s,m_b^2,m_u^2)}{s}
\right)\right.
\nonumber\\
&&+\left. \frac{1}{pq_1}\left(
\frac{\lambda^{1/2}(s,m_b^2,m_u^2)}{2s}-
m_1^2\log\left(\frac{s+m_1^2-m_2^2+\lambda^{1/2}(s,m_1^2,m_2^2)}
{s+m_1^2-m_2^2-\lambda^{1/2}(s,m_1^2,m_2^2)}\right)\right)
\right\}.  
\end{eqnarray}
For $q_1^2=0$ one has $pq_1=(M_B^2-M_\rho^2)/2$. 

Now, let us analyze the behaviour of the form factor in the limit $m_b\to\infty$. 
To this end it is convenient to rewrite the spectral representation (\ref{fadisp}) 
in terms of the light-cone variables as follows (see \cite{lc} for details) 
\begin{eqnarray}
\label{fa-lc}
\frac{2}{M_B}F_{A}^{(1)}(m_1,m_2)&=&\frac{\sqrt{N_c}}{4\pi^2}\int \frac{dx_1 dx_2 dk_\perp^2}{x_1^2 x_2}\delta(1-x_1-x_2)
\frac{\phi_B(s)}{s-M_\rho^2}\left\{m_1x_2+m_2 x_1-(m_1-m_2)k_\perp^2/pq_1\right\}.  
\end{eqnarray}
Here $x_i$ is the fraction of the $B$-meson light-cone momentum carried by the 
quark $i$, and $s=m_1^2/x_1+m_2^2/x_2+k_\perp^2/x_1x_2$. 
For the form factors $F_{A}^{(u)}$ and $F_{A}^{(b)}$ one obtains 
\begin{eqnarray}
\label{fau}
\frac{2}{M_B}F_{A}^{(u)}&=&-Q_u\frac{\sqrt{N_c}}{4\pi^2}\int \frac{dx dk_\perp^2}{x_u^2 x_b}
\frac{\phi_B(s)}{s-M_\rho^2}\left\{m_u\,x_b+m_b\,x_u+\frac{2(m_b-m_u)k_\perp^2}{M_B^2-M_\rho^2}
\right\}, \nonumber
\\
\frac{2}{M_B}F_{A}^{(b)}&=&\;\;\;Q_b\frac{\sqrt{N_c}}{4\pi^2}\int \frac{dx dk_\perp^2}{x_b^2 x_u}
\frac{\phi_B(s)}{s-M_\rho^2}\left\{m_b\,x_u+m_u\,x_b+\frac{2(m_u-m_b)k_\perp^2}{M_B^2-M_\rho^2}
\right\}, \nonumber
\end{eqnarray}
with 
\begin{eqnarray}
s=\frac{m_b^2}{x_b}+\frac{m_u^2}{x_u}+\frac{k_\perp^2}{x_u x_b}.     
\end{eqnarray}
Let us recall that the $B$-meson decay constant has the following representation 
in terms of the wave function \cite{m}: 
\begin{eqnarray}
\label{fp}
f_B&=&\frac{\sqrt{N_c}}{4\pi^2}\int \frac{dx dk_\perp^2}{x_u x_b}
\frac{\phi_B(s)}{s-M_\rho^2}\left\{m_u\,x_b+m_b\,x_u\right\}. 
\end{eqnarray}
Due to the wave function $\phi_B(s)$, the integral in the heavy quark limit is dominated 
by the region $x_u=\bar\Lambda/m_b$, $x_b=1-\bar\Lambda/m_b$, 
where $\bar\Lambda$ is a constant of order $M_B-m_b$. This leads to the following expansion 
of the form factors in the $1/m_b$ series 
\begin{eqnarray}
\label{hqlimit}
\frac{2}{M_B}F_{A}^{(u)}&=&-Q_u\frac{f_b}{\bar\Lambda m_b}+...\nonumber\\
\frac{2}{M_B}F_{A}^{(b)}&=&\;\;\,\,Q_b\frac{f_b}{m_b^2}+...
\end{eqnarray}
Clearly, the dominant contribution in the heavy quark limit comes from the process 
when the light quark emits the photon, whereas the emission of the photon from the heavy 
quark gives only a $1/m_b$ correction. The expressions (\ref{hqlimit}) for the form factor 
$F_{A}^{(u)}$ agrees with the result of Ref. \cite{korch}, 
while we have found a different sign for $F_{A}^{(b)}$. 

\subsection{The form factor $F_V$} 
\begin{center}
\begin{figure}[hb]
\mbox{\epsfig{file=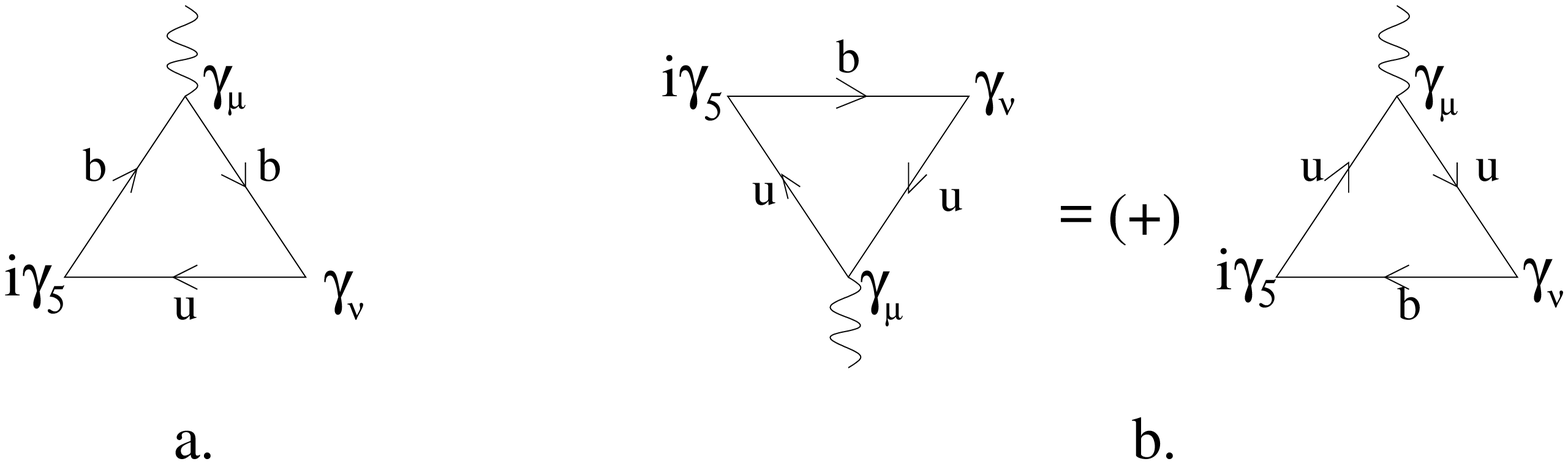,width=10cm}}
\caption{\label{fig:Fv}Diagrams for the form factor $F_V$: a) $F_V^{(b)}$, b) $F_V^{(u)}$.} 
\end{figure}
\end{center}

The consideration of the form factor $F_{V}$ is very similar to the form factor $F_{A}$. 
$F_V$ is determined by the two diagrams shown in Fig \ref{fig:Fv}:  
Fig \ref{fig:Fv}a gives $F_V^{(b)}$, the contribution of the process when the $b$ quark interacts with the 
photon; Fig \ref{fig:Fv}b describes the contribution of the process when the quark $u$ interacts. 
One has    
\begin{eqnarray}
F_V=F_V^{(b)}+F_V^{(u)}. 
\end{eqnarray}
It is again convenient to change the direction of the quark line in the loop diagram of 
Fig \ref{fig:Fv}b 
describing $F_V^{(u)}$ 
by performing the charge conjugation of the matrix element. For the vector current $\gamma_\nu$ 
in the vertex the sign does not change (in contrast to the $\gamma_\nu\gamma_5$ case 
considered above). 

\begin{center}
\begin{figure}[hb]
\mbox{\epsfig{file=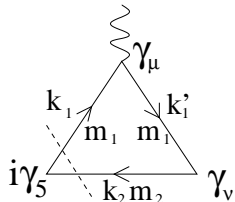,width=3cm}}
\caption{\label{fig:Fvt} The triangle diagram for $F_V^{(1)}(m_1,m_2)$. 
The cut corresponds to calculating the imaginary part in the variable $p^2$.}
\end{figure}
\end{center}

Then both diagrams in Fig \ref{fig:Fv} a, b are 
reduced to the diagram of Fig \ref{fig:Fvt} which gives the 
form factor $F_V^{(1)}(m_1,m_2)$: Setting $m_1=m_b$, $m_2=m_u$ gives $F_V^{(b)}$: 
$F_V^{(b)}=Q_b F_V^{(1)}(m_b,m_u)$; Setting $m_1=m_u$, $m_2=m_b$ 
gives $F_V^{(u)}$, $F_V^{(u)}=Q_u F_V^{(1)}(m_u,m_b)$. 
The trace corresponding to the diagram of Fig 4 (1 - active quark, 2 - spectator) reads 
\begin{eqnarray}
-{\rm Sp}\;
i\gamma_5(m_2-\hat k_2)\gamma_\nu (m_1+\hat k'_1)\gamma_\mu(m_1+\hat k_1)=
-4\epsilon_{\nu\mu \alpha q_1}(m_1 k_2+m_2 k_1)_\alpha. 
\nonumber
\end{eqnarray}
The spectral representation for the form factor takes the form
\begin{eqnarray}
\nonumber
\frac{2}{M_B}F_V^{(1)}=\frac{\sqrt{N_c}}{4\pi^2}\int\limits_{(m_b+m_u)^2}^\infty
\frac{ds\phi_B(s)}{(s-M_\rho^2)}\left\{
(m_2-m_1)\frac{\lambda^{1/2}(s,m_b^2,m_u^2)}{s}+
m_1\log\left(\frac{s+m_1^2-m_2^2+\lambda^{1/2}(s,m_1^2,m_2^2)}
{s+m_1^2-m_2^2-\lambda^{1/2}(s,m_1^2,m_2^2)}\right)
\right\}. 
\end{eqnarray}
To analyze the heavy quark limit $m_b\to\infty$ we again represent the form factor 
in terms of the light-cone variables 
\begin{eqnarray}
\label{fv-lc}
\frac{2}{M_B}F_{V}^{(1)}&=&-\frac{\sqrt{N_c}}{4\pi^2}
\int \frac{dx_1 dx_2 dk_\perp^2}{x_1^2 x_2}\delta(1-x_1-x_2)
\frac{\phi_B(s)}{s-M_\rho^2}\left(m_1x_2+m_2 x_1\right).  
\end{eqnarray}
For $F_{V}^{(u)}$ and $F_{V}^{(b)}$ the corresponding expressions read 
\begin{eqnarray}
\label{fvu}
\frac{2}{M_B}F_{V}^{(u)}&=&-Q_u\frac{\sqrt{N_c}}{4\pi^2}\int \frac{dx dk_\perp^2}{x_u^2 x_b}
\frac{\phi_B(s)}{s-M_\rho^2}\left\{m_u\,x_b+m_b\,x_u\right\}, \nonumber
\\
\frac{2}{M_B}F_{V}^{(b)}&=&-Q_b\frac{\sqrt{N_c}}{4\pi^2}\int \frac{dx dk_\perp^2}{x_b^2 x_u}
\frac{\phi_B(s)}{s-M_\rho^2}\left\{m_b\,x_u+m_u\,x_b\right\}, \nonumber
\end{eqnarray}
By the same procedure as used for $F_A$, we obtain in the limit $m_b\to\infty$ 
\begin{eqnarray}
\label{hqlimit1}
\frac{2}{M_B}F_{V}^{(u)}&=&-Q_u\frac{f_b}{\bar\Lambda m_b}+...\nonumber\\
\frac{2}{M_B}F_{V}^{(b)}&=&-Q_b\frac{f_b}{m_b^2}+...
\end{eqnarray}
The dominant contribution in the heavy quark limit again comes from the process 
when the light quark emits the photon. Now both form factors $F_{V}^{(u)}$ and $F_{V}^{(b)}$
in (\ref{hqlimit1}) perfectly agree with the expansions obtained in \cite{korch}.  

As seen from Eqs. (\ref{hqlimit}) and (\ref{hqlimit1}), one finds $F_A=F_V$ in the heavy quark limit, 
in agreement with the large-energy effective theory \cite{leet}. 

\section{The form factor $G_V$}
\begin{center}
\begin{figure}[hb]
\mbox{\epsfig{file=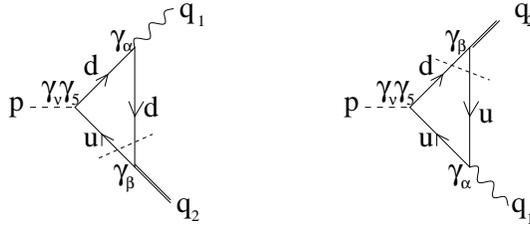,width=7cm}}
\caption{\label{fig:rhogamma} Diagrams describing the amplitude 
$\langle \gamma(q_1)\rho^-(q_2)|\bar d \gamma_\nu\gamma_5 q|0\rangle$, $p=q_1+q_2$, $p^2=M_B^2$. 
The diagram (a) is multiplied by the $d$-quark charge, and the 
the diagram (b) is multiplied by the $u$-quark charge. 
The cut corresponds to the calculation of the imaginary part in the variable $q_2^2$. }
\end{figure}
\end{center}

The form factor $G_V$ is determined by the divergence of the matrix element of the charged current 
between the vacuum and the $\rho^-\gamma$ state,
\begin{eqnarray}
ip_\nu\langle \gamma(q_1)\rho^-(q_2)|\bar d\gamma_\nu\gamma_5 u|0\rangle=
eG_V \epsilon_{q_1\epsilon_1^\ast q_2\epsilon_2^\ast}. 
\end{eqnarray}
The corresponding diagrams are shown in Fig \ref{fig:rhogamma}.  
If the classical equations of motion are applied, the form factor $G_V$ is proportional to the 
light-quark masses.   
This is the reason why this form factor was neglected in previous 
analyses \cite{gp,sr1}. However, a proper calculation shows that this argument is not correct: 
in fact the classical equations 
of motions do not hold and the divergence contains the anomaly. 

The anomalous behavior of the divergence of the axial-vector current in the chiral limit 
is a well-known phenomenon discovered in the two-photon amplitude 
$\langle \gamma\gamma|\partial_\nu \bar q\gamma_\nu\gamma_5 q|0\rangle$ \cite{abj}. 
A very clear way to demonstrate the anomaly is to start with the matrix element of the axial vector 
current and to calculate the spectral representations for the relevant form factors. 
The anomaly is then obtained by performing the divergence at the final stage of the calculation \cite{dz}.  
A similar treatment applied to the matrix element $\langle \gamma\rho^-|\bar d\gamma_\nu\gamma_5 u|0\rangle$
leads to the form factor $G_V$ which does not vanish in the limit $m\to 0$. 
We will present a detailed discussion of the anomaly in radiative $B$ decays in a separate 
publication \cite{ms1}. Here we only quote the final result: considering the spectral 
representation in the variable $q_2^2$ and introducing the appropriate $\rho$-meson radial wave function 
$\psi_\rho(s)$, one finds the following expression for the form factor $G_V$ 
\begin{eqnarray}
\label{gvd}
G_V=\sqrt{N_c}(Q_u+Q_d)\left[-\frac{M_B^2}{4\pi^2}\int\frac{ds\;\psi_\rho(s)\,(s-M_\rho^2)}{(s-M_B^2-i0)^2}\right].  
\end{eqnarray} 
$\psi_\rho(s)$ is normalized according to the relation (for massless quarks) 
\begin{eqnarray}
\label{norma1}
\frac{1}{8\pi^2}\int\limits_{0}^\infty ds\; s\;|\psi_\rho(s)|^2=1.  
\end{eqnarray}
Clearly, $G_V$ is finite for $m=0$ which means a violation of the classical equations
of motion for the axial-vector current. Eq. (\ref{gvd}) describes the anomaly which 
takes place for the $\gamma\rho$ final state as well as for the $\gamma\gamma$ one. 
There is however an important difference between the two cases: 
For the $\gamma\gamma$ final state the divergence remains finite in the limit $m\to 0$ and 
$p^2=M_B^2\to\infty$. 
For the $\rho\gamma$ final state the divergence is finite for $m\to 0$ but 
decreases as $1/p^2$ for $p^2\to\infty$. 
It is convenient to introduce the parameter $\kappa$ such that 
\begin{eqnarray}
\label{kappa}
G_V= (Q_u+Q_d)\;\kappa\frac{M_\rho f_\rho}{M_B^2},  
\end{eqnarray} 
with $\kappa$ staying finite for $m=0$ and $M_B\to\infty$.

\section{Numerical estimates}
Let us write once more the penguin and the weak annihilation contributions to 
the $B^-\to\rho^-\gamma$ amplitude: 
\begin{eqnarray}
F^{\rm peng}_{\rm PV}&=&F^{\rm peng}_{\rm PC}=-\xi_t C_7\frac{m_b}{2\pi^2}T_1(0),
\nonumber\\
F^{\rm WA}_{\rm PV}&=&\xi_u a_1 M_\rho f_\rho\;\left[\frac{2F_{A}}{M_B}+\frac{2f_B}{M_B^2-M_\rho^2} \right].
\nonumber\\
\label{formulas}
F^{\rm WA}_{\rm PC}&=&\xi_u a_1 M_\rho f_\rho\;\left[\frac{2F_V}{M_B}-\frac{f_B G_V}{M_\rho f_\rho}\right]. 
\end{eqnarray}
The scaling behavior of the form factors in the limit $M_B\to\infty$ reads 
\begin{eqnarray}
T_1(0)\sim M_B^{-3/2} \cite{scaling}, 
\quad F_{A,B}\sim M_B^{-1/2},  \quad G_V\sim M_B^{-2} 
\end{eqnarray}
such that 
\begin{eqnarray}
F^{\rm peng}\sim M_B^{-1/2}, \quad F^{\rm WA}_{\rm PV,PC}\sim M_B^{-3/2}. 
\end{eqnarray}
The terms proportional to $f_B$ ($f_B\sim 1/\sqrt{M_B}$) are $1/M_B$-suppressed 
compared with the terms containing $F_{A,V}$. 
As we see below numerically this leads to a suppression by a factor of $4-5$. 

We now proceed to numerical estimates for the $B$-meson decay. 
The scale-dependent Wilson coefficients $C_i(\mu)$ and $a_1(\mu)$ take the following values
at the renormalization scale $\mu\simeq 5$ GeV \cite{gsw}: 
\begin{eqnarray}
C_1=1.1,\qquad  C_2=-0.241, \qquad  C_{7\gamma}=-0.312, \qquad a_1=C_1+C_2/N_c\simeq 1.02.   
\end{eqnarray}
The penguin form factor was previously calculated within the dispersion approach with the result 
$T_1^{B^-\to\rho^-}(0)=0.27\pm 0.3$ \cite{ms}. Using the same parameters and the $B$ meson wave 
function as in \cite{ms} we obtain the form factors $F_{A,B}$ shown in Table I. 
Our result for the form factor $F_V$ is in good agreement with the estimates from other approaches. 
The form factor $F_A$ agrees well with the constraints from perturbative QCD and turns out to be considerably 
larger than the corresponding sum rule estimate. 

The value of the $G_V$ is very sensitive to the details of the $\rho$ meson wave function  
$\psi_\rho$. The reason for that is the presence of the term $(s-M_\rho^2)$ in the integrand in Eq. 
(\ref{gvd}) which changes sign in the integration region. Assuming $\psi_\rho(s)\simeq {\beta^2}/{(\beta^2+s)^2}$ and setting 
$\beta=0.8$ GeV, which gives a good description of the $\rho$ meson radius, leads to 
$\kappa=-1.8$. However, this value has a large uncertainty.  
Conservatively, we take $|\kappa|<2.0$ and use this result for further estimates. 
The form factor $G_V$ does not contribute more than a few \% to the full amplitude, but 
can sizeably correct the weak-annihilation part. 

Using the obtained form factors and the decay constants $f_B=0.18$ GeV, $f_\rho=0.21$ GeV 
we arrive at the following estimates (we show contributions of different terms in Eq. (\ref{formulas}) separately):  
\begin{eqnarray}
F^{\rm peng}&=&\xi_t\;20\;{\rm MeV},  
\nonumber\\
F^{\rm WA}_{\rm PV}&=&\xi_u\; \left\{
-7.8\;({\rm cont.\; of}\; F_A)+\; 
 2.2\;({\rm cont.\; of}\; f_B)\right\}\;{\rm MeV}  
=-5.6\;\xi_u\;{\rm MeV}
\nonumber\\
F^{\rm WA}_{\rm PC}&=&\xi_u\;\left\{-6.0\;({\rm cont.\; of\;} F_V)
\pm\;0.7\rm ({cont.\; of\;} G_V)\right\}\;{\rm MeV}
=-(6.0\pm 0.7)\;\xi_u\;{\rm MeV}.
\end{eqnarray}

Taking into account errors in the form factors, for the ratios of the 
weak-annihilation to the penguin amplitudes we find  
\begin{eqnarray}
\label{results}
{\rm parity-violating}  &:&\qquad {F^{\rm WA}_{\rm PV}}/{F^{\rm peng}_{\rm PV}}
\simeq -(0.28\pm 0.025)\;{\xi_u}/{\xi_t}, 
\nonumber\\
{\rm parity-conserving} &:&\qquad {F^{\rm WA}_{\rm PC}}/{F^{\rm peng}_{\rm PC}}
\simeq -(0.3\pm 0.05)\;{\xi_u}/{\xi_t}. 
\end{eqnarray}
$|\xi_u/\xi_t|\simeq 0.4$ in the Standard Model. 

Our results for $F^{\rm peng}$ and $F^{\rm WA}$ agree with the sum rules 
\cite{sr2} which reported the ratio  
$F^{\rm WA}/F^{\rm peng}\simeq -0.3\;{\xi_u}/{\xi_t}$.  
We would like to notice, however, that the anatomy of $F^{\rm WA}$ in our analysis is 
different. For instance, we have found $F_A$ considerably larger than the sum rule result. 
But after including the contact term $\sim f_B$ which was not taken into account in 
\cite{sr2}, we have come to $F^{\rm WA}_{\rm PV}$ close to the sum rule estimate. 

The calculated form factors and the ratios of the weak-annihilation to the penguin
amplitudes is one of the necessary ingredients for the calculation of the 
branching ratios of the $B\to\rho\gamma$ decays, Isospin and CP Asymmetries. 
In addition to the above ratios, these quantities contain the phase induced by the strong 
interactions and the CP-violating phase of the CKM matrix (see \cite{ap,ppp} for details). 
The corresponding analysis was done recently in \cite{ap} using the value  
$F^{\rm WA}/F^{\rm peng}\simeq -0.3\;{\xi_u}/{\xi_t}$.

\begin{table}[htb]
\caption{\label{table:results} Results for the weak-annihilation form factors 
$F_A$, $F_V$ and $G_V$ for the $B^-\to\rho^-\gamma$ decay. The accuracy of our estimates 
is about 10\%. The sum rule results are recalculated from \protect\cite{sr2} according to the relation 
$F_{A,V}=-F^{\rm SR}_{1,2}{M_B}/{f_{\rho^-}}$. The results from \protect\cite{korch} 
are recalculated according to $F_{A,V}=-\frac{1}{2}f^{\protect\cite{korch}}_{A,V}$ for 
$\bar \Lambda=0.5$ GeV.}
\centering
\begin{tabular}{|l|r|r|r|}
    & Disp Approach & SR \cite{sr2}  &  pQCD\cite{korch}  \\
\hline
$-F_{A}$   & 0.120   & 0.073   & $>$0.09   \\
$-F_{V}$   & 0.092   & 0.091   & $>$0.09   \\
$|G_V|$    & $<$0.004  &         &                
\end{tabular}
\end{table}

\section{Conclusion}
We have analyzed the weak annihilation for the radiative decay $B\to\rho\gamma$ 
in the factorization approximation. 

\noindent 1. We have calculated the form factors $F_A$ and $F_V$ describing   
the photon emission from the $B$ meson loop within the relativistic dispersion 
approach. We have performed the $1/m_b$ expansion of the form factors and demonstrated that the form factors 
of the dispersion approach exhibit a behaviour in agreement with the large-energy limit of QCD. 

\noindent 2. We have analyzed the contribution to the weak annihilation amplitude from the diagram 
when the photon is emitted from the loop containing only light quarks. 
For the parity-conserving process this quantity is related to the divergence of 
the axial-vector current 
\begin{eqnarray}
\langle \gamma(q_1)\rho^-(q_2)|\partial_\nu \bar d \gamma_\nu \gamma_5 u|0\rangle=
e\epsilon_{q_1\epsilon_1^\ast q_2\epsilon^\ast_2}(Q_u+Q_d)\;\kappa\;{f_\rho M_\rho}/{M_B^2}. 
\end{eqnarray}
with $\kappa$ staying finite in the chiral limit $m\to 0$. 
This result means the violation of the classical equations of motion and represents an anomaly which has the 
same origin as the anomaly of the matrix element  
$\langle \gamma(q_1)\gamma(q_2)|\partial_\mu A_\mu|0\rangle$. 
The value of $\kappa$ is found to be sensitive to subtle details of the $\rho$-meson wave function.  
Conservatively, we estimate $|\kappa|\lesssim 1$. 

\noindent 3. We have also included contact terms which were missed in some of the 
previous analyses. Numerical estimates for the 
weak annihilation contribution to the $B^-\to\rho^-\gamma$ amplitude are given 
in Eq. (\ref{results}).  

It was noticed in \cite{sr1} that the weak annihilation mechanism
is crucial for the $D\to \rho\gamma$ decays in which it dominates over the penguin mechanism.  
It is therefore very important to take into account all the contribitions to the weak 
annihilation amplitude listed above for a proper description of the rare radiative $D$ decays. 

\acknowledgments 
We are grateful to O. Nachtmann, O. P\`ene and 
V. I. Zakharov for useful discussions and to K. Schubert for interest in this work.  
We also like to thank Th. Feldmann for a critical reading of a preliminary 
version of the paper. The work was supported by the Alexander von Humboldt Stiftung, 
BMBF project 05 HT 9 HVA3, and the Deutsche Akademische Austauschdienst (DAAD).

\section{Appendix: transversity of the parity-violating amplitude. }
\noindent The parity-violating amplitude (\ref{apva}) is given by the sum of the two terms 
\begin{eqnarray}
A^{\rm PV}(B^-\to\rho\gamma^-)\sim A^{\rm PV}_1+A^{\rm PV}_2  
\end{eqnarray}
where 
$A^{\rm PV}_1=\langle \rho^-(q_2)|\bar d \gamma_\nu u|0 \rangle 
\langle \gamma(q_1)|\bar u \gamma_\nu\gamma_5 b|B^-(p) \rangle$ 
and $A^{\rm PV}_2=
\langle \rho^-(q_2)\gamma(q_1)|\bar d \gamma_\nu u|0 \rangle 
\langle 0|\bar u \gamma_\nu\gamma_5 b|B^-(p) \rangle$. 

\noindent 1. We start with $A^{\rm PV}_1$. Let us write 
$\langle \gamma(q_1)|\bar u \gamma_\nu\gamma_5 b|B^-(p) \rangle=e\,\epsilon_1^{\ast\mu} T^B_{\mu\nu}$ with 
\begin{eqnarray}
T^B_{\mu\nu}(p,q)&=&i\int dx e^{iqx}\langle 0|T(J_\mu(x),\bar u\gamma_\nu \gamma_5 b)|B^-(p)\rangle,
\end{eqnarray}
where 
$J_\mu(x)=\frac23\bar u\gamma_\mu u
-\frac13 \bar d\gamma_\mu d
-\frac13 \bar b\gamma_\mu b$ is the electromagnetic quark current. 
The amplitude has the following Lorentz structure 
\begin{eqnarray}
\label{qq}
T^B_{\mu\nu}=i\left(g_{\mu\nu}pq_1-p_{\mu}q_{1\nu}\right)F_{1A}(q_1^2)
+i\left(q_1^2\,p_{\mu}-{pq_1}\,q_{1\mu}\right)q_{1\nu}F_{2A}(q_1^2)
+i\left(q_1^2\,p_{\mu}-{pq_1}\,q_{1\mu}\right)p_{\nu}F_{3A}(q_1^2)
+\frac{ip_{\mu}p_\nu}{pq_1} C, 
\end{eqnarray}
where $C$ is the contact term. 
The contact term can be determined using the conservation of the electromagnetic current 
$\partial_\mu J_\mu=0$, which leads to the relation 
\begin{eqnarray}
\label{eq3}
q_\mu T^B_{\mu\nu}(p,q)=-\langle 0|[\hat Q, \bar d\gamma_\nu \gamma_5 u]|B^-(p)\rangle
=Q_Bf_B p_\nu=-if_B p_\nu  
\end{eqnarray}
for the $B^-$ meson. This gives $C=-f_B$. Notice that for the $B^0$-meson the contact term is absent. 
For the radiative decay $q_1^2=0$ so we find 
\begin{eqnarray}
A^{\rm PV}_1&=&
ie\,f_\rho M_\rho \epsilon_1^{\ast\mu}\epsilon_2^{\ast\nu}\left\{
(g_{\mu\nu}pq_1-p_{\mu}q_{1\nu})F_{1A}(0)-p_\mu p_{\nu} \frac{2f_B}{M_B^2-M_\rho^2}\right\}
\nonumber\\
&=&
ie\,f_\rho M_\rho \epsilon_1^{\ast\mu}\epsilon_2^{\ast\nu}\left\{
(g_{\mu\nu}pq_1-p_{\mu}q_{1\nu})F_{1A}(0)-p_\mu q_{1\nu}\frac{2f_B}{M_B^2-M_\rho^2}\right\}. 
\end{eqnarray}

\noindent 2. Using the equation of motion for the quark fields ($Q_u=2/3\;e$, $Q_d=-1/3\;e$)
\begin{eqnarray}
i\gamma_\nu\partial^\nu q(x)&=& m q(x) - Q_q A_\nu \gamma^\nu q(x), 
\nonumber\\
i\partial^\nu \bar q(x)\gamma_\nu&=& -m \bar q(x) + Q_q A_\nu \bar q(x)\gamma^\nu , 
\end{eqnarray}
one obtains for $A_2^{\rm PV}$ 
\begin{eqnarray}
A^{\rm PV}_2&=&ip_\nu f_B \langle \rho^-\gamma|\bar d \gamma_\nu u|0 \rangle=
f_B \langle \rho^-\gamma|\partial_\nu(\bar d \gamma_\nu u)|0 \rangle
\nonumber\\
&=&-(Q_d-Q_u)e\,\langle \rho^-\gamma|A^\nu \bar d \gamma_\nu u|0 \rangle
=e\,\epsilon_1^{\ast\mu}\epsilon_2^{\ast\nu} g_{\mu\nu}f_\rho M_\rho f_B +O(m_u,m_d). 
\end{eqnarray}  

\noindent 3. For the sum we find 
\begin{eqnarray}
A^{\rm PV}_1+A^{\rm PV}_2=ie\,f_\rho f_B M_\rho \epsilon_1^{\ast\mu}\epsilon_2^{\ast\nu}
\left(g_{\mu\nu}pq_1-p_{\mu}q_{1\nu}\right)\left[F_{1A}(0)+\frac{2f_B}{M_B^2-M_\rho^2}\right]
\end{eqnarray}  
and obtain the relation (\ref{fpv}) after setting $F_{1A}=2F_A/M_B$. 
For the $B^0\to \rho^0\gamma$ decay the term proportional to $f_B$ is absent. 

\end{document}